# On The Significance of Which-Way Expositions: Propounding a New Possibility

Mohammad Bahrami[1], Afshin Shafiee[2]

Department of Chemistry, Sharif University of Technology

P.O.Box, 11365-9516, Tehran, Iran

**Abstract**

In this article, we survey some controversial problems concerning the idea of erasing Which Way information proposed in recent years. A statistical examination of these proposals suggests that whenever the Bayesian rule is taken into account for two relevant events in two successive times, the probabilistic description of them is unavoidably time-symmetric. Consequently, it seems that they cannot fulfill the implications of a so-called delayed-choice experiment. As a possible alternative, however, we suggest a new experimental arrangement in which one can change the whole state of a given system at a proper time (without measurement) to accomplish an actual delayed-choice experiment with a time-asymmetric attribute. The peculiar features of this experiment are then discussed.

**Keywords**

Which Way Experiments, Quantum Eraser, Complementarity Principle, Delayed-Choice, Physical Reality.

# 1 Introduction

If we consider *interference* and *entanglement* as the most distinctive *mysteries* of quantum world, then Which Way (WW) experiments will be at the *heart of quantum experiments* which *dramatically* illustrate the differences between classical and quantum *conceptions* of nature.

---

[1] mbahrami@mehr.sharif.edu
[2] Corresponding author: shafiee@sharif.edu

The idea of WW experiments could be traced back to the famous double-slit experiment with movable slit proposed by Einstein to show the inconsistency of orthodox interpretation of quantum theory [1]. After that, in his response to the so-called EPR paradox, Bohr gave an account of this experiment based on his interpretation of quantum mechanics [2]. Later, Wootter and Zureck investigated this problem by quantum mathematical formalism [3]. In 1982, the idea of Quantum Eraser (QE) of WW Information (WWI) was proposed by Scully and Drühl [4], and later in 1991 it was examined again by Scully, Englert and Walther (SEW) with the aid of Micromasers as WW-marker [5]. Since then, all the conceptual and experimental problems concerning the idea of erasing WWI have been analyzed theoretically or experimentally in light of different viewpoints [6-18].

By obtaining WWI or the existence of detectable WWI (that makes the paths to be distinguishable), the interference pattern will be destroyed. Then, what will happen if WWI is removed (erased)? Could the interference pattern be restored? What if the WWI is erased after the detection of atoms on the detector-plate in double-slit experimental arrangement? By delaying the erasure process of WWI, does *observer's choice* affect the interference pattern? Does delayed-choice erasing of *past* information really change the *present* observed information? Generally, these are the main conceptual questions about WW experiments and the idea of WW Information Erasing (WWIE).

In the following, we are going to examine the key feature of a WWIE processes by focusing on SEW proposal as a representative portrayal. We will argue that the results of SEW could be consistently explained in a coherent statistical approach without reference to any possible interpretation of quantum mechanics. Our examination shows that the whole description in these so-called delayed-choice experiments is statistically time-symmetric. So, we are in doubt whether these schemes could be really suggestive. This is our main business in sections 2 and 3. Subsequently, in section 4, we propose a simple WWIE experiment which satisfies the implications of a delayed-choice experiment. In section 5, we discuss about two possible WWIE proposals including a measurement process and a non-unitary action without measurement. We show that our proposed experiment is of the second type. Then, we argue that the reality assumption has a prominent role in explaining peculiar features of a double-slit experiment such as ours. Section 6 includes some concluding remarks.

## 2  Which Way experiments

Here we are focusing on the main scheme of SEW article [5] which is conceptually a *representative account* of the other similar works [6-18]. In this context, our first assumption is that the WWI is stored somewhere *rather than* the interfering quantum system. Accordingly, the observer could erase the WWI whenever he wants to do so, at any desired time.

Now, let us consider the experimental setup proposed by SEW in Fig.1. The cavities in this figure are resonant systems in which a suitable excited atom emits a photon with probability equal to one when travelling through it without significant disturbance on the atoms' de Broglie wave length. The same probability is zero when the atom is outside of the cavities[1]. The composite system of cavities is the physical system in which the WWI is stored and the interfering quantum system is a suitable atom.

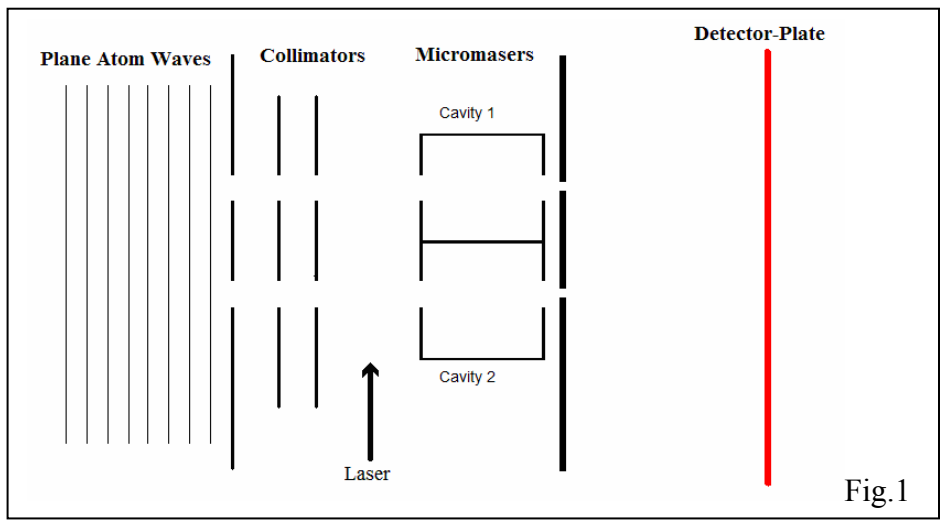

Fig.1

When the laser is turned off, the state function of atoms in the region between the slits and the detector-plate is described by

$$\Psi(\vec{r}) = \frac{1}{\sqrt{2}}[\psi_1(\vec{r}) + \psi_2(\vec{r})]|g\rangle_{Atom} \quad (1)$$

where $\vec{r}$ denotes the centre-of-mass coordinate of a given atom, $|g\rangle_{Atom}$ is the ground-internal state of atoms and $\psi_j(\vec{r})$ is diffracted wave function originating when slit $j$ is open ($j = 1, 2$). Regarding (1), the probability of finding atoms on the detector-plate at $\vec{r} = \vec{R}$ is given by

$$P(\vec{R}) = |\Psi(\vec{R})|^2 = \frac{1}{2}\left[|\psi_1(\vec{R})|^2 + |\psi_2(\vec{R})|^2 + 2\operatorname{Re}(\psi_1^*(\vec{R})\psi_2(\vec{R}))\right] \quad (2)$$

where $\operatorname{Re}(\psi_1^*(\vec{R})\psi_2(\vec{R}))$ is the interference term. In fact, (2) demonstrates the pattern obtained by the data observed on the detector-plate when the whole system is prepared in the state (1). For our next purposes, we call this situation as (I).

---

[1] More detailed analysis of micromasers could be found in [19].

Next, consider the situation in which the laser in turned on. Then, the state function of atoms and cavities in the region between the slits and the detector-plate is given by

$$\Psi(\vec{r}) = \frac{1}{\sqrt{2}}\left[\psi_1(\vec{r})|1_1 0_2\rangle_{cavities} + \psi_2(\vec{r})|0_1 1_2\rangle_{cavities}\right]|g\rangle_{Atom} \quad (3)$$

where $|1_1 0_2\rangle_{cavities}$ describes the situation in which one photon is in cavity-1 and none in cavity-2, and $|0_1 1_2\rangle_{cavities}$ describes the situation in which one photon is in the cavity-2 while there is none in cavity-1. The reader should keep in mind that both $|1_1 0_2\rangle_{cavities}$ and $|0_1 1_2\rangle_{cavities}$ are physical states of the composite system containing two cavities (two sub-system). These states include the accessible WWI. Since the atoms can pass through just one of the cavities and not both, and meanwhile the $|1_1 0_2\rangle_{cavities}$ & $|0_1 1_2\rangle_{cavities}$ are distinguishable physical states of composite system of the cavities, then it is apparent that $\langle 1_1 0_2 | 0_1 1_2 \rangle = 0$.

If the observer just analyzed the detector-plate data, then the probability of finding atoms at $\vec{r} = \vec{R}$ on the detector-plate would be

$$P(\vec{R}) = |\Psi(\vec{R})|^2 = \frac{1}{2}\left\{\begin{array}{l} |\psi_1(\vec{R})|^2 + |\psi_2(\vec{R})|^2 + \psi_1^*(\vec{R})\psi_2(\vec{R})\langle 1_1 0_2 | 0_1 1_2 \rangle \\ + \psi_1(\vec{R})\psi_2^*(\vec{R})\langle 0_1 1_2 | 1_1 0_2 \rangle \end{array}\right\}$$

$$= \frac{1}{2}\left(|\psi_1(\vec{R})|^2 + |\psi_2(\vec{R})|^2\right) \quad (4)$$

As is obvious in (4), there would be no interference pattern in this new situation although no WWI is observed. We call this situation as (II,a).

If the experimenter observes the existence of photon in each cavity, then there could be two different situations:

1. Finding the photon at cavity 1. Here the observed information includes the presence of a photon in cavity-1 as well as the location of the atoms at $\vec{r} = \vec{R}$ on the detector-plate. Consequently, one observes the space distribution of atoms as $\frac{1}{2}|\psi_1(\vec{R})|^2$.

2. Finding the photon at cavity 2. Here the statistics is built on the presence of a photon in cavity-2 as well as the location of the atoms at $\vec{r} = \vec{R}$ on the detector-plate. Consequently, one observes the space distribution of atoms as $\frac{1}{2}|\psi_2(\vec{R})|^2$.

We call the above situations as (II,b) and (II,c) respectively.

If in any way the cross terms in relation (4) (including the inner products $\langle 1_1 0_2 | 0_1 1_2 \rangle$ and $\langle 0_1 1_2 | 1_1 0_2 \rangle$) do not vanish, then the interference pattern will be reappeared. The complete recovery of interference pattern is the main purpose of a WWIE process. Nevertheless, there is no doubt that the proper interactions for WWIE cannot be described by unitary transformations, because for any unitary transformation we should have

$$\langle (1_1 0_2)' | (0_1 1_2)' \rangle = \langle 1_1 0_2 | U^\dagger U | 0_1 1_2 \rangle = \langle 1_1 0_2 | 0_1 1_2 \rangle$$

which entails that the probability of (4) should not be changed. To observe the effect of a WWIE process, a non-unitary change of state (e.g., a state-reduction) should take place. In other words, a desired erasing process should necessarily include a non-unitary change of state for which a quantum measurement is a natural alternative. To our best knowledge, this is the common feature of all WWIE proposals that their erasing process is a quantum *measurement* [6-18].

SEW proposed an experimental setup depicted in Fig.2 in which the further possibility of doing an erasure process (i.e. a proper quantum measurement on photon in the cavities) is controlled by two shutters placed between the cavities and the Detector Wall (D.WW). SEW envisaged D.WW as an atom with a ground state *g* and an excited state *e*.

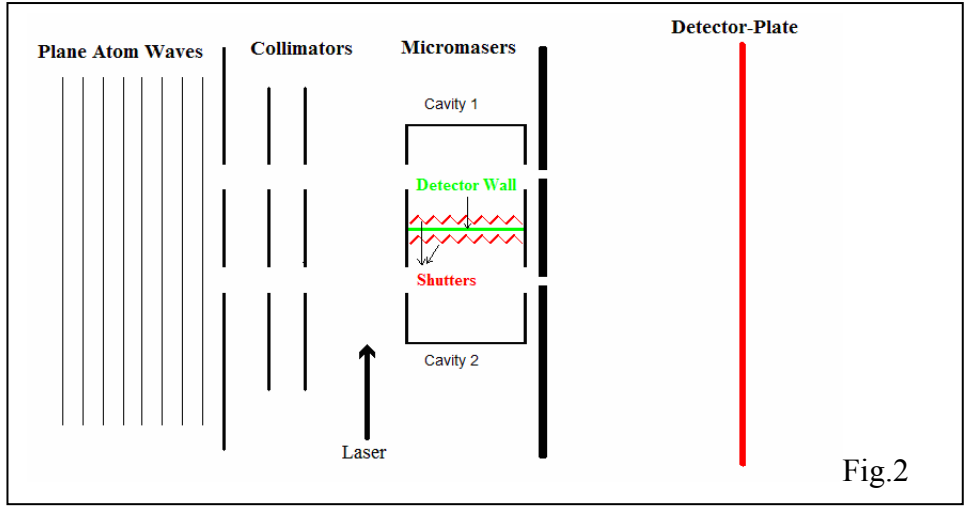

Fig.2

If the shutters are closed while the atoms are passing through the cavities and the slits, the state function of whole system (including the atom, D.WW and cavities) will be described as:

$$\Psi(\vec{r}) = \frac{1}{\sqrt{2}} \left[ \psi_1(\vec{r}) |1_1 0_2\rangle_{cavities} + \psi_2(\vec{r}) |0_1 1_2\rangle_{cavities} \right] |g\rangle_{Atom} |g\rangle_{Detector} \qquad (5)$$

where $|g\rangle_{Detector}$ is the ground state of D.WW. One can rewrite the above relation as:

$$\Psi(\vec{r}) = \frac{1}{\sqrt{2}} \left[ \psi_+(\vec{r}) |+\rangle_{cavities} + \psi_-(\vec{r}) |-\rangle_{cavities} \right] |g\rangle_{Atom} |g\rangle_{Detector} \qquad (6)$$

where $|\pm\rangle_{\text{cavities}} = \frac{1}{\sqrt{2}}\left(|1_1 0_2\rangle_{\text{cavities}} \pm |0_1 1_2\rangle_{\text{cavities}}\right)$ and $|\pm\rangle_{\text{cavities}} = \frac{1}{\sqrt{2}}\left(|1_1 0_2\rangle_{\text{cavities}} \pm |0_1 1_2\rangle_{\text{cavities}}\right)$.

The relation (6) is more convenient for describing the interaction between the cavities fields and the D.WW. Considering the interaction Hamiltonian between the cavities and D.WW, only $|+\rangle_{\text{cavities}}$ (which denotes a symmetric state) will be coupled to the cavities fields. So when the shutters are opened and the D.WW interacts with cavities, the state function of (6) becomes

$$\Psi(\vec{r}) = \frac{1}{\sqrt{2}}\left[\psi_+(\vec{r})|0_1 0_2\rangle_{\text{cavities}}|e\rangle_{\text{Detector}} + \psi_-(\vec{r})|-\rangle_{\text{cavities}}|g\rangle_{\text{Detector}}\right]|g\rangle_{\text{Atom}} \quad (7)$$

where $|e\rangle_{\text{Detector}}$ is the excited state of the D.WW.

Here if an experimenter only takes into account the data of the detector-plate, the obtained pattern will be

$$P(\vec{R}) = |\Psi(\vec{R})|^2 = \frac{1}{2}\left[|\psi_+(\vec{R})|^2 + |\psi_-(\vec{R})|^2\right]$$
$$= \frac{1}{2}\left[|\psi_1(\vec{R})|^2 + |\psi_2(\vec{R})|^2\right] \quad (8)$$

which is the same as (4), except for the fact that there is no WWI here. We call this situation as (III,a).

But if the experimenter considers the data collected at the detector-plate and the status of D.WW both together, he will get

$$P(\vec{R}) = |\psi_+(\vec{R})|^2 = \frac{1}{4}\left[|\psi_1(\vec{R})|^2 + |\psi_2(\vec{R})|^2 + 2\text{Re}\left(\psi_1^*(\vec{R})\psi_2(\vec{R})\right)\right] \quad (9)$$

Where (9) denotes the probability of finding atoms at $\vec{r} = \vec{R}$ on the detector-plate when the D.WW is in the excited-state. Similarly the probability of finding atoms at $\vec{r} = \vec{R}$ on the detector-plate when D.WW is in the ground-state is given by

$$P(\vec{R}) = |\psi_-(\vec{R})|^2 = \frac{1}{4}\left[|\psi_1(\vec{R})|^2 + |\psi_2(\vec{R})|^2 - 2\text{Re}(\psi_1^*(\vec{R})\psi_2(\vec{R}))\right] \quad (10)$$

As is obvious in relations (9) and (10), the interference term is appeared again. We call the situations described by (9) and (10) as (III,b) and (III,c), respectively. It is also clear that $(8) = (9) + (10)$. Here, the important point is that when the shutters are opened and the D.WW interacts with cavities, after the measurement being completed, the photon state will collapse into $|+\rangle_{\text{cavities}}$ or $|-\rangle_{\text{cavities}}$. These new states possess no WWI (i.e., the probability for each path is the same in the either $|+\rangle_{\text{cavities}}$ or $|-\rangle_{\text{cavities}}$).

We can summarize all the above possible situations as the following:

I. The laser is turned off. Then the observed data are only the ones collected on the detector-plate which show a complete interference pattern.

II. The laser is turned on and the shutters are closed. Then:

(a) By analyzing the data on the detector-plate, the observed pattern will be $\frac{1}{2}\left[|\psi_1(\vec{R})|^2 + |\psi_2(\vec{R})|^2\right]$.

(b) If one finds a photon in cavity-1 and observes the location of atoms on the detector-plate, the observed pattern will be $\frac{1}{2}|\psi_1(\vec{R})|^2$.

(c) If one finds a photon in cavity-2 and observes the location of atoms on the detector-plate, the observed pattern would be $\frac{1}{2}|\psi_2(\vec{R})|^2$.

III. The laser is turned on and the shutters are opened. Then:

(a) By analyzing the data on the detector-plate, the observed pattern will be $\frac{1}{2}\left[|\psi_1(\vec{R})|^2 + |\psi_2(\vec{R})|^2\right]$.

(b) By observing the location of atoms at the Detector-Plate and finding the D.WW in the excited state, the obtained pattern will be

$$\frac{1}{4}\left[|\psi_1(\vec{R})|^2 + |\psi_2(\vec{R})|^2 + 2\text{Re}(\psi_1^*(\vec{R})\psi_2(\vec{R}))\right].$$

(c) By observing the location of atoms at the Detector-Plate and finding the D.WW in the ground state, the obtained pattern will be

$$\frac{1}{4}\left[|\psi_1(\vec{R})|^2 + |\psi_2(\vec{R})|^2 - 2\text{Re}(\psi_1^*(\vec{R})\psi_2(\vec{R}))\right].$$

## 3  Physical Interpretation, Complementarity and Delayed-Choice

All situations mentioned above are manifestations of the *Complementarity Principle*; they are *complementary phenomena* under *different experimental conditions*. In each *well defined* experimental setup that enables us to find *exact* information about one of the system's degrees of freedom, we lose (the possibility of obtaining) exact information about other complementary degrees of freedom. Any complete information about a dynamical variable of system precludes the complete information about some other complementary one. This is the quintessential and distinctive feature of *quantum phenomena*. In this way, the situations (I) and (II), as well as (II,b,c) and (III,b,c) are complementary: in situation (I) the wave pattern, in (II) the particle pattern and in (III,b,c) again the wave-pattern are observed.

To give a clear explanation of the Complementarity in the above situations, we define some relevant observables as the following:

- $\vec{R}_0$ is the position of atoms on the Detector-Plate, corresponding to the $\hat{X}_{\text{atoms}}$ position operator satisfying the relation $\hat{X}\phi_{R_0} = \vec{R}_0 \phi_{R_0}$, where $\phi_{R_0}$ is the eigenfunction of $\hat{X}_{\text{atoms}}$. $X_{\text{atoms}}$ is the corresponding observable.

- The observation of photons in each one of the cavities is described by $|1_1\rangle$ or $|1_2\rangle$ [1], corresponding to the $\hat{\sigma}_{\|(\text{cav})}$ operator in relations $\hat{\sigma}_{\|(\text{cav})}|1_1\rangle = +|1_1\rangle$ and $\hat{\sigma}_{\|(cav)}|1_2\rangle = -|1_2\rangle$. We assign a "+" ("-") value whenever we find a photon in cavity 1(2). $\sigma_{\|(cav)}$ is the corresponding observable.

---

[1] Here we define $|1_2 0_2\rangle_{\text{cavities}}$ as $|1_1\rangle_{\text{cav}}$ and $|0_1 1_2\rangle_{\text{cavities}}$ as $|1_2\rangle_{\text{cav}}$.

- The observation of D.WW in the excited or ground states is corresponded with the observing of photons in the states $|+\rangle_{cavities}$ or $|-\rangle_{cavities}$ at the cavities, respectively. Here we define a $\hat{\sigma}_{\perp(cav)}$ operator which its eigenstates are $|\pm\rangle$ where $\hat{\sigma}_{\perp(cav)}|\pm\rangle = \pm|\pm\rangle$. The corresponding observable is $\sigma_{\perp(cav)}$. The eigenvalues of "+" or "-" are related to the status of the D.WW in the excited or ground state, respectively. Regarding the relations of $|\pm\rangle = \frac{1}{\sqrt{2}}[|1_1\rangle \pm |1_2\rangle]$, $\hat{\sigma}_{\perp(cav)}$ and $\hat{\sigma}_{\|(cav)}$ operate like the spin Pauli operators $\hat{\sigma}_x$ and $\hat{\sigma}_z$ of a spin-half system, respectively.

Now, we consider a composite system including the atoms and the cavities, so that the situations (II) and (III) are about some given measurements on this composite system. In the situations (II,b,c), the corresponding observables of $\hat{X}_{atoms} \otimes \hat{\sigma}_{\|(cav)}$ are measured, and in (III,b,c) the corresponding observables of $\hat{X}_{atoms} \otimes \hat{\sigma}_{\perp(cav)}$. Since $[\hat{\sigma}_{\|(cav)}, \hat{\sigma}_{\perp(cav)}] \neq 0$, one can also show that $[\hat{X}_{atoms} \otimes \hat{\sigma}_{\|(cav)}, \hat{X}_{atoms} \otimes \hat{\sigma}_{\perp(cav)}] \neq 0$. Thus the situations of (II,b,c) and (III,b,c) are complementary.

Regarding the incompatibility of $\sigma_{\|(cav)}$ and $\sigma_{\perp(cav)}$, the preparation of the system in an eigenstate of $\sigma_{\|(cav)}$ (which describes the situation when the shutters are closed and there is one photon in one of the cavities) means that we have exact information about $\sigma_{\|(cav)}$. Then, for such a system, after the measurement of $\sigma_{\perp(cav)}$, we will lose accessible exact knowledge about $\sigma_{\|(cav)}$. In other words, knowing any value of $\sigma_{\perp(cav)}$ will result in a complete loss of information about $\sigma_{\|(cav)}$. This is exactly what we mean by saying that *the $\sigma_{\perp(cav)}$ measurement is a WWIE process*. Here the idea of WWIE is directly resulted from the incompatibility of $\sigma_{\|(cav)}$ and $\sigma_{\perp(cav)}$.

The situations (III,b,c) describe a joint measurement of $X_{atoms}$ and $\sigma_{\perp(cav)}$. Regarding the time sequence of measurements, this can be achieved in two different procedures:

- **(a1)** $X_{atoms}$ is first measured and then $\sigma_{\perp(cav)}$.
- **(a2)** $\sigma_{\perp(cav)}$ is first measured and then $X_{atoms}$.

Although the above circumstances have a time order, they could be statistically described by the *same* joint probability. To show this, we denote the joint probability of events *A* and *B* (*A* and *B* are *compatible* events) under a given specific condition *C* (determined by the experimental arrangements) as $p(A, B | C)$. Here, we characterize *A*, *B* and *C* as

- *C* : the preparation of the composite system of atom and cavities in state (5) or (6), which is determined by an appropriate experimental arrangement.
- *A* : the observation of atoms at $\vec{R}_0$ on the Detector-Plate which corresponds to a measurement of $X_{atoms}$.
- *B* : the observation of D.WW in the ground or excited state which corresponds to a measurement of $\sigma_{\perp(cav)}$.

In (III,b,c), the experimenter measures both the $X_{atoms}$ and $\sigma_{\perp(cav)}$ whose results are statistically described by $p(A, B | C)$. To obtain $p(A, B | C)$, the observer can choose one of the **a1** or **a2** procedures:

- In **a1**, he first obtains $p(A|C)$ and then $p(B|A,C)$. So he can measure $p(A, B | C)$ as $p(A, B | C) = p(A | C) p(B | A, C)$.
- In **a2**, he first obtains $p(B|C)$ and then $p(A|B,C)$. So he can measure $p(A, B | C)$ as $p(A, B | C) = p(B | C) p(A | B, C)$.

As a result, independent of any physical interpretation, the time sequence of our observations on *A* or *B* has no influence on calculating $p(A, B | C)$. The probabilistic description of events is time-symmetric here. This is a consequence of the Bayesian rule for the joint probability of two events in different occasions when there is no preferred order of time in statistical characterization of events. So it can be easily concluded that in the above mentioned situations, the delayed mode of measurement has no special importance.

Meanwhile, we suggest that the time-independent behavior of all above situations (particularly (III)) needs no further interpretation. All of the *minimalistic* knowledge-based interpretation [8], the *reality-of-phenomena* interpretation [9], and the *Bohmian* interpretation [18] approve the symmetric behavior of joint probability $p(A,B|C)$ for two *compatible* events *A* and *B* in two different times. So, we conclude that these situations cannot fulfill the implications of a delayed-choice experiment in which the time sequence of events should have a direct effect on the final result [20,21].

In addition, for $p(A|C)$ we have:

$$p\left(X_{atoms} = \vec{R}_0 | C\right) = p\left(X_{atoms} = \vec{R}_0 \ \& \ \sigma_{\perp(cav)} = +1 | C\right)$$
$$+ p\left(X_{atoms} = \vec{R}_0 \ \& \ \sigma_{\perp(cav)} = -1 | C\right) \quad (11)$$

The relation (11) gives us the same result as $(8) = (9) + (10)$. In addition, the data required for calculating $p\left(X_{atoms} = \vec{R}_0 \ \& \ \sigma_{\perp(cav)} = \pm 1 | C\right)$ is undoubtedly the sub-ensemble of the data ensemble required for calculation of $p\left(X_{atoms} = \vec{R}_0 | C\right)$. We should also notice that only $p\left(X_{atoms} = \vec{R}_0 \ \& \ \sigma_{\perp(cav)} = \pm 1 | C\right)$ shows the interference pattern and not $p\left(X_{atoms} = \vec{R}_0 | C\right)$. In other word, the interference pattern is just recovered for *sub-ensembles* of the whole ensemble of the data obtained on the detector-plate.

In effect, we can summarize our results in the following clause:

- The probabilistic description of SEW proposal is time-symmetric. This is not in accordance with the concept of a delayed-choice experiment in which the time order of events should have a prominent role in its quantum description.

Yet, in the following, we are going to describe a new WWIE setup which has distinctive traits in comparison with the pervious works. In our proposed experiment, the time order of erasing process and the detection event has a direct effect on the final pattern. Thus the idea of delayed-choice is strictly demonstrated here.

# 4  Erasing the possibility of obtaining WW Information

In this section, we are going to propose a WWIE protocol in which the key point is that the erasing process is not a quantum measurement but still includes a non-unitary change of state. Let us consider again the experimental setup of SEW in Fig.3.

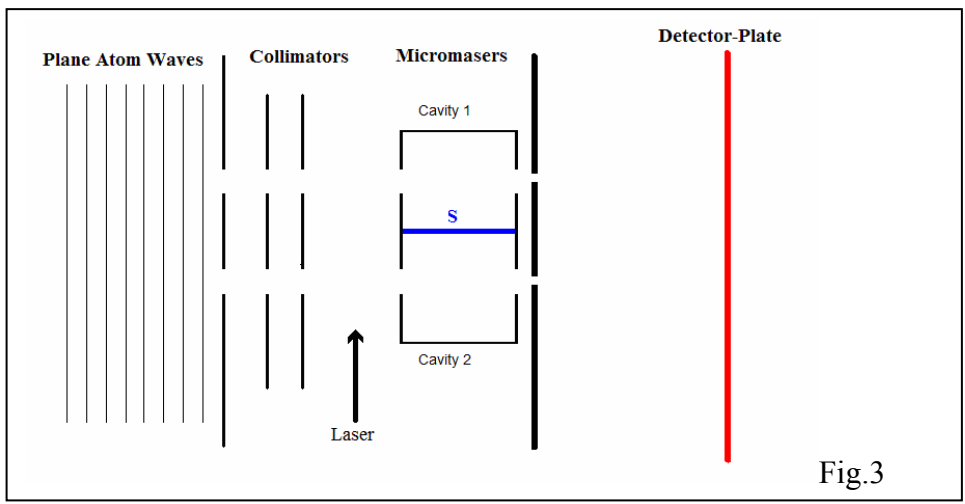

Fig.3

Here the plate S that separates two cavities is removable. When S is placed, the state function of atoms and cavities is given by

$$\Psi(\vec{r}) = \frac{1}{\sqrt{2}}\left[\psi_1(\vec{r})|1_1 0_2\rangle_{cavities} + \psi_2(\vec{r})|0_1 1_2\rangle_{cavities}\right]|g\rangle_{Atom} \quad (12)$$

which is the same as (3). So the final pattern on the detector-plate is predicted by the following relation:

$$P(\vec{R}) = |\Psi(\vec{R})|^2 = \frac{1}{2}\left\{\begin{array}{l}|\psi_1(\vec{R})|^2 + |\psi_2(\vec{R})|^2 + \psi_1^*(\vec{R})\psi_2(\vec{R})\langle 1_1 0_2 | 0_1 1_2\rangle \\ + \psi_1(\vec{R})\psi_2^*(\vec{R})\langle 0_1 1_2 | 1_1 0_2\rangle\end{array}\right\}$$

$$= \frac{1}{2}\left(|\psi_1(\vec{R})|^2 + |\psi_2(\vec{R})|^2\right) \tag{13}$$

which is the same as the relation (4).

Now suppose that when the atoms are still flying between the slits and the detector-plate, we remove the plate S, so that, the state of atoms and cavities after removing of S can be given by

$$\Psi(\vec{r}) = \frac{1}{\sqrt{2}}\left[\psi_1(\vec{r})|1\rangle_{\text{cavity T}} + \psi_2(\vec{r})|1\rangle_{\text{cavity T}}\right]|g\rangle_{\text{Atom}}$$

$$= \frac{1}{\sqrt{2}}\left[\psi_1(\vec{r}) + \psi_2(\vec{r})\right]|1\rangle_{\text{cavity T}} |g\rangle_{\text{Atom}} \tag{14}$$

where $|1\rangle_{\text{cavity T}}$ means that one photon exists in the whole cavity (cavity T) at each run of the experiment. Then, the observed pattern on the detector-plate is given by

$$P(\vec{R}) = |\Psi(\vec{R})|^2 = \frac{1}{2}\left\{|\psi_1(\vec{R})|^2 + |\psi_2(\vec{R})|^2 + 2\,\text{Re}\left(\psi_1^*(\vec{R})\psi_2(\vec{R})\right)\right\} \tag{15}$$

We call this situation as (IV,a).

Here, the composite system of cavities (containing two sub-systems) is changed into a simple one. The reader should carefully distinguish between this new situation and the previously mentioned situations (III,b,c). In the latter situations, the two cavities are still distinguishable even while the shutters are being opened, because there still exist D.WW which differentiates between two cavities. By removing the differentiating plate (S), the distinguishable states of cavities (which describe a composite system) are changed into one unique state $|1\rangle_{\text{cavity T}}$ (which describes a simple system). In other word, when S is removed, the distinguishable paths are changed into the indistinguishable ones. Therefore, one cannot obtain the WWI, although the photon is not erased too. In fact, in our proposed experiment, the *possibility* of obtaining WWI is erased.

Although the change of the state from (12) to the (14) is not caused by a measurement process, it is still a non-unitary change of state, because the probabilities in (13) and (15) are different. So it will fulfill the requirements of an erasing process. We have accomplished this kind of change by removing the plate S at a given moment.

If S was removed after the detection of atoms on the detector-plate, no interference pattern would be obtained (although the WWI is erased) and the observer could not distinguish ways the atoms had passed. Here the erasure of WWI gives no new information for the observer. In this situation, the mere information we have is the location of atoms' center-of-mass on the detector-plate (i.e., the detection of $X_{atoms}$) which was observed under the state preparation described by (12). Thus the observed pattern would be $\frac{1}{2}\left[|\psi_1(\vec{R})|^2 + |\psi_2(\vec{R})|^2\right]$, the same result as (13). We call this situation as (IV,b). One should notice here that although the observed patterns in (II,a) and (IV,b) are the same, in (II,a) the observer could still distinguish the way that atoms passed, but in (IV,b) there is no such a possibility.

Furthermore, the time of erasing plays a key role here: to observe the interference pattern the experimenter should erase the possibility of obtaining WWI only during the time interval when atoms are between the slits and the detector-plate . In contrast to (III) in which the erasing process had no observable effect on the statistical results of the location of atoms (i.e., the values of $X_{atoms}$), here in (IV), during a specific time interval, the erasing process has a direct effect on the type of pattern we anticipate to observe.

Here again, the probability theory gives a clear and helpful insight to the problem. Let us define the following events:

- $C$: the preparation of both the atoms and the measuring instrument in state (5) or (6).
- $C'$: the preparation of both the atoms and the measuring instrument in state (14).

- $A$: the detection of atoms at $\vec{R}$ on the Detector-Plate $(X_{atoms} = \vec{R})$.

In the situations (IV,a) and (IV,b), what the observer measures is $p(A|C)$ and $p(A|C')$ respectively. After measuring the location of atoms on the detector-plate, however, the change of conditions has no effect on the observed data. This is exactly in accordance with the predictions of quantum mechanics. Our supposed time interval is also the time in which the measurement has not yet performed and so any change in the conditions will definitively change the results of the latter measurements. But after the detection of atoms on the detector-plate, any change in conditions will have no effect anymore.

It is also clear that $p(X_{atoms} = \vec{R}|C')$ (which signifies the interference pattern) is obtained by the whole ensemble of the data observed on the detector-plate. In other word, the interference pattern is recovered for the whole ensemble of data observed on the detector-plate, but not sub-ensembles of it.

## 5 Discussion

The WWIE proposals could be classified in two main different classes:

1. WWIE1: The erasing process is a non-unitary quantum measurement. The WWI is erased by a measurement of an observable.
2. WWIE2: The erasing process is different from a measurement but still caused by a non-unitary action. Here no new information is obtained by erasure.

Both WWIE1 and WWIE2 are *manifestations* of the Complementarity Principle, but the probabilistic description of WWIE2 is time-asymmetric. Our proposed experiment is a WWIE2 type. As Bohr carefully insisted, in quantum realm, any change in *experimental arrangement* could result in a new *phenomenon*, i.e. some new possible observable results. It is exactly what we see in (IV,a,b) situations.

In Bohr's point of view, the new phenomenon is characterized by some new possible *uncontrollable and unavoidable* **interactions** with the measuring agency. But in our proposed experiment, the erasure can be done when the atoms are sufficiently far away from the cavities, where no known local interactions could be considered acting between the atoms and the cavities. In addition, it is clear that the information stored in cavities is information about the *past* history of atoms, including the path that atoms passed through.

In contrast to quantum physics, we have no witness of the delayed-choice effect in classical physics. The classical entities are assumed to be real in the various circumstances. For example a ball cannot be *blue* and *yellow* simultaneously, but this dual character does not bring about Complementarity (in its quantum sense) in the macro world. In order to determine a clear and correct physical interpretation, consider the following examination, illustrated in Fig.4.

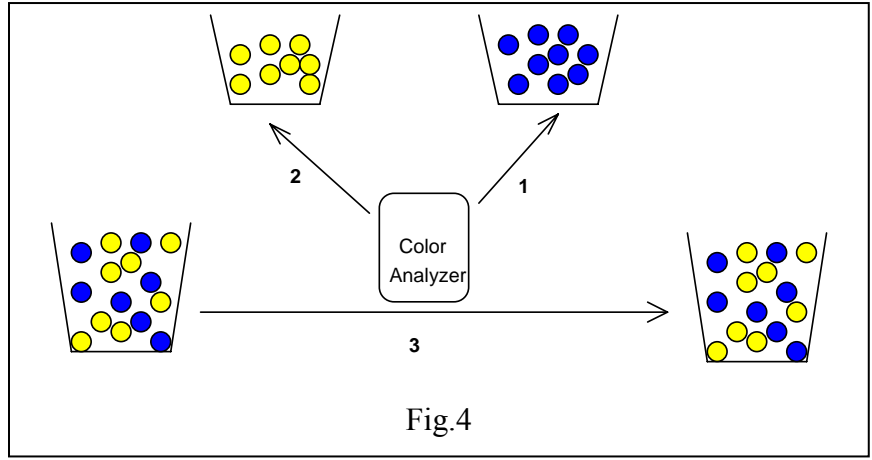

Fig.4

Let us assume that we are going to separate some balls according to their colors by a suitable color analyzer. With a color analyzer present, we assume that the balls go through one of the channels "1" or "2", and in the absence of it, they go through the channel "3". What will happen, however, if we remove (or destroy) the color analyzer after the balls have passed through it? Apparently no change will happen! Because here the balls are physical entities whose reality is independent of the way we arrange to see them.

Whenever one neglects the role of reality assumption in quantum realm, many misleading attitudes and misconceptions may arise. In a realistic perspective, however, the state function of quantum mechanics might not completely describe the physical reality[1]. So, the controversial problems, such as the delayed-choice effect, are then interpreted subordinately. For example, in Bohm's approach [23], some of the bizarre aspects of micro-world are resolved by considering the real character of particles guided (and influenced) by waves through a quantum potential, however, the problem of non-locality is raised here. In Bohr's point of view, on the other hand, it is impossible to describe the whole reality of phenomena and this impossibility is a peculiar characteristic of micro physics, claimed to be expressed by Complementarity Principle [1,2]. Bell also showed the problem is much more fundamentally challenging to be solved [24], but not impossible. There are still some other ways going beyond the limitations of Bell's inequalities [25,26].

## 6   Conclusion

Reviewing the well-known representation of SEW by demonstration a WWIE process at different situations, we argued that the significance of a delayed-choice experiment could not be illustrated in such expositions. Then we suggested a new experimental arrangement in which the WWI could be erased without measurement. The WWIE process, here, depends on what the observer decides to perform before the particle's detection. This is a delayed-choice experiment, since the time order of events has a prominent role in its quantum (probabilistic) description. Our proposed experiment shows that the reality assumption is decisive in expounding the peculiar features of a double-slit experiment and its corresponding process.

## 7   Acknowledgment

One of the authors (M. Bahrami) would like to appreciate Časlav Brukner and Hans Leydolt for their valuable discussions on an earlier version of this paper.

---

[1] This is the stand point of Einstein, Podolsky and Rosen (EPR) in their famous article in 1935 [22].